\documentclass[12pt]{article}

\def \eqtext#1		{\hspace{1in} \hbox{#1}}	
\providecommand{\micron}{\ensuremath{\mu\rm{m}}}

\def \etal      	\hbox{ \it et al.} 		
\def \etals		\hbox{{\it et al.}'s}		
\def \vs		\hbox{vs.}			
\providecommand{\degrees}{\ensuremath{{}^\circ}}

\def \idlplot#1		{\centerline{\scalebox{0.9}{\includegraphics{#1}}}} 
\def \idlplotps#1	{\centerline{\scalebox{0.9}{\includegraphics[70,350][574,710]{#1}}}} 

\def \eg		{{\it e.g.\/}}
\def \subsimt#1{{\lower 2pt\hbox{$\scriptstyle #1$}\atop
     \raise 1pt\hbox{$\scriptstyle \sim$}}}

\def \lesssim    	{\subsimt <}

\newcommand{\promille}{ 
  \relax\ifmmode\promillezeichen
        \else\leavevmode\(\mathsurround=0pt\promillezeichen\)\fi}
\newcommand{\promillezeichen}{%
  \kern-.05em%
  \raise.5ex\hbox{\the\scriptfont0 0}%
  \kern-.15em/\kern-.15em%
  \lower.25ex\hbox{\the\scriptfont0 00}}

\newcommand{\ltsim}	{\ensuremath{\lesssim}}

 1
 2

 1
 2

\usepackage{natbib}
\usepackage{graphics}
\usepackage{graphicx}
\usepackage{wrapfig}
\usepackage{tocloft}
\usepackage[outercaption]{sidecap}  

\usepackage{mdwlist}

\begin{document}

\renewcommand{\textfraction}{0.15}              
\renewcommand{\bottomfraction}{0.65}
\renewcommand{\floatpagefraction}{0.60}

\bibliographystyle{icarus}

\markright{Throop: NH Ring Collision Hazard, v2}
\pagestyle{myheadings}
\thispagestyle{empty}				

\title{New Horizons Ring Collision Hazard: \\
Constraints from Earth-based Observations \\
\ \\
A White Paper for the New Horizons Mission
\ \\ 
}

\author{Henry Throop\\
Planetary Science Institute\\
\texttt{throop@psi.edu}}

\date{Version 1: 21-Dec 2011\\
Version 2: 31-Jan-2013\\
\  \\
\textit{Version 2 fixes an error in how diffraction was handled in reflected light, and improves the scattering model by
considering non-Mie particles.}}

\maketitle




\begin{abstract}
The New Horizons spacecraft's nominal trajectory crosses the planet's satellite plane at $\sim 10,000\ \rm{km}$ from the
barycenter, between the orbits of Pluto and Charon.  I have investigated the risk to the spacecraft based on
observational limits of rings and dust within this region, assuming various particle size distributions.  The best
limits are placed by 2011 and 2012 HST observations, which significantly improve on the limits from stellar occultations,
although they do not go as close to the planet.  From the HST data and assuming a `reasonable worst case' for the size
distribution, we place a limit of $N < 20$ damaging impacts by grains of radius $> 0.2\ \textrm{mm}$ onto the spacecraft
during the encounter.  The number of hits is $\approx$ 200$\times$ above the NH mission requirement, and $\approx$
$2000\times$ above the mission's desired level. Stellar occultations remain valuable because they are able to measure
$N$ closer to the Pluto surface than direct imaging, although with a sensitivity limit several orders of magnitude
higher than that from HST imaging. Neither HST nor occultations are sensitive enough to place limits on $N$ at or below
the mission requirements.  \end{abstract}

\section{Background and Motivation}

On November 3-4, 2011, the ``New Horizons Encounter Hazards Workshop" workshop was held at Southwest Research Institute,
Boulder, CO. The purpose of the workshop was to discuss possible collisional hazards to New Horizons (NH) during its
upcoming 2015 trajectory through the Pluto system. Discussion topics included a variety of theoretical modeling and
observations, each of which placed limits on a range of particle sizes at different locations within the system.  Based
on these inputs and followup work, the NH mission will decide on a final trajectory at some point prior to the 2015
encounter. This paper summarizes and follows up on work I presented at the workshop.

\section{Mission Requirements}

The New Horizons spacecraft incorporates a multi-layer flexible shield, which in part protects the spacecraft against
damage from high-speed micrometeoroid dust impacts.  The shield is designed to protect NH against impacts of $3 \times
10^{-4} \ \rm{g}$ at relevant speeds. The New Horizons mission itself has defined a slightly lower critical mass of $m_c
= 1\times10^{-4}\ \rm{g}$.  The mission has a requirement of $N < 0.1$ impacts of size $m_c$ or greater during the
encounter, and a `desirement' of $N <0.01$ impacts of mass $m_c$. For the purpose of this paper, I assume a density of
4~g~cm$^{-3}$, corresponding to a critical radius $r_c = 0.2\ \textrm{mm}$.  



\section{Observational Limits}

\textbf{Rings in reflected light.} Observations by HST and ground-based telescopes can be used to search directly for
rings and dust orbiting in the region near Pluto. Assuming a non-resolved uniform ring of optical depth~$\tau<1$, the
observed brightness in reflected light is

\begin{equation}
{I \over F} = {\tau a \over \mu} {P(\alpha) \over P(0\degrees)}
\label{eq:iof}
\end{equation}

In this expression, $I$ is the measured ring intensity and $\pi F$ is the incident solar flux. The ring's optical depth
is $\tau$, and albedo is $a$. $\mu$ is the cosine of the angle $B$, which is the tilt angle of the rings (from normal)
as seen by the observer.  $P(\alpha)$ is the particle's phase function at scattering angle $\alpha$.  For Earth-based
observations of Pluto, $\alpha$ is in the range 0--2$\degrees$.

The normal optical depth for a ring seen in reflected light is

\begin{equation}
\tau(\lambda) = \mu {I\over F}(\lambda) = \int n(r)\, \pi r^2\, Q_{sca}(r,\lambda)\, P(\alpha)dr  ,
\label{eq:tau}
\end{equation}
where $n(r) dr$ is the size distribution of the dust grains, with units $\# \rm{cm}^{-2}$.
The scattering efficiency $Q_{sca}$ is usually computed using Mie scattering, and approaches the value of $1$ for $r \gg
\lambda$ and moderately absorbing particles. 



Finally, the number of damaging hits to the spacecraft during its passage through the
system can be written as

\begin{equation}
N = \int n(r > r_c)\, {A\, \over \mu_{sc}} dr
\label{eq:ndamage}
\end{equation}
where $A\approx 5\ \rm{m}^2$ is the cross-sectional area of the spacecraft (Chris Hersman, personal communication),
and $\mu_{sc}$ is the cosine of the sub-spacecraft latitude $B_{sc}$ at closest approach. A trajectory directly through
the plane will encounter fewer particles than one on a shallow slant angle; for the nominal NH encounter geometry,
GeoViz\footnote{\texttt{http://soc.boulder.swri.edu/nhgv}} shows that $B_{sc} \approx 49\degrees$.  Combining Eqs.
\ref{eq:iof}--\ref{eq:ndamage}, $N$ can be calculated exactly from $I/F$ and the size distribution $n(r)$.

\textbf{Rings in extincted light.} In the case of extincted light (\eg, a stellar occultation), the ring's brightness is
calculated slightly differently. Here, the observed intensity $I$ is

\begin{equation}
I = I_0\, e^{-\tau/\mu}\ ,
\end{equation}
where $I_0$ is the unocculted stellar intensity. The optical depth is given by
\begin{equation}
\tau(\lambda) = \int n(r)\, \pi r^2\, Q_{ext}(r,\lambda) \ dr .
\label{eq:tauoccult}
\end{equation}
where $Q_{ext}$ is the Mie extinction efficiency. Eqs. \ref{eq:ndamage}--\ref{eq:tauoccult} can then be used to calculate
$N$ from $\tau$ and $n(r)$. Note that in extinction, the particle albedo is not relevant.




\section{Ring Upper Limits from Stellar Occultations}

Stellar occultations have two distinct advantages over direct imaging. First, they offer a higher resolution $\delta R$,
defined by the Fresnel limit

\begin{equation} 
\delta R = \sqrt(30\ \rm{AU}\, \lambda/2) \approx 2\ \rm{km},
\end{equation}
rather than the diffraction limit, which in the case of HST is 
\begin{equation}
\delta R = (30\ \rm{AU})\ 1.22\, \lambda / (2.4\ \rm{m}) \approx 1500\ \rm{km} .
\end{equation}

Second, the occultation observations are immune to the effects of stray light. This allows for measurements to continue
with no loss in sensitivity even as the star and occulting body approach each other on the detector. Imaging experiments
(like on HST) are often limited by stray light, in particular when searching for faint features such as rings at
sub-arcsecond separations from a bright extended object such as Pluto.

\begin{figure}
\centerline{\scalebox{0.6}{\includegraphics{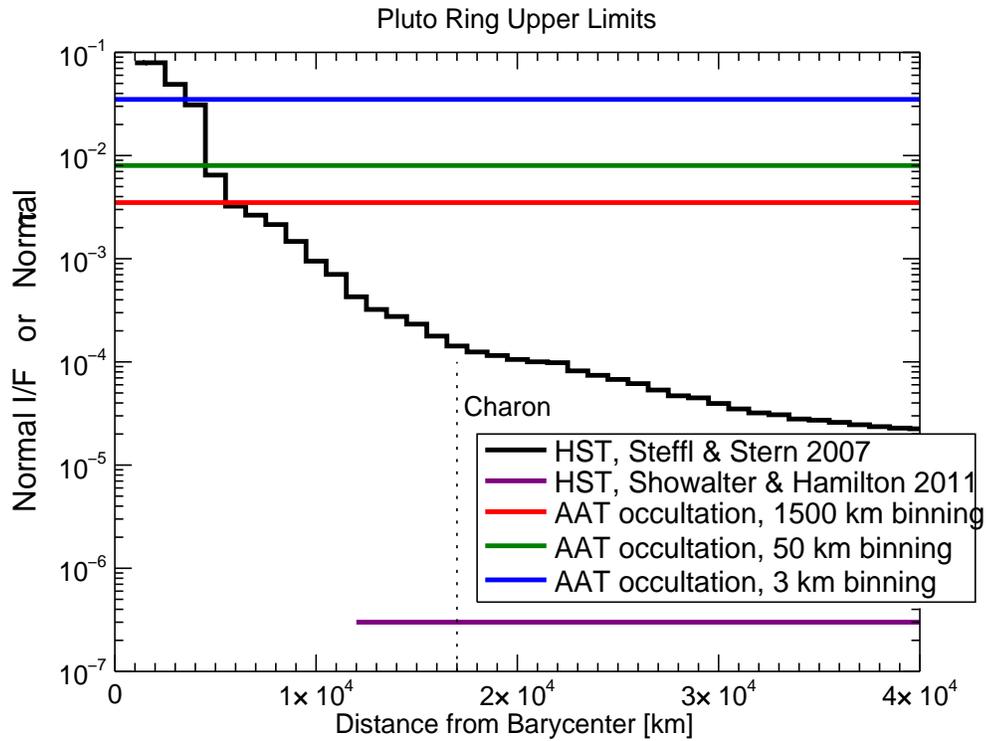}}}
\caption{Observational upper limits for rings in the Pluto system. The 2007 HST observations were affected by stray light
which increased moving closer to Pluto. The 2011 HST observations reduced stray light significantly but were saturated
inward of 12,000~km. HST's spatial resolution was 1500~km. The 2006 AAT
stellar occultation is unaffected by stray light, and provides the best limit near the system barycenter.
2007 HST data provided by \hbox{A.} Steffl and adapted from \citet{ss07}.}
\label{fig:tau}
\end{figure}

\textbf{2006 AAT occultation.} Pluto passed in front of the magnitude 15.5 star P384.2 on 2006 Jun 12.  This occultation
was observed by R.~French and K.~Shoemaker using the 3.9-meter Anglo-Australian Telescope (AAT). This observation coupled a
large aperture with a low-noise, fast-readout CCD (10 fps) to study the system at unprecedented resolution. The data
achieved a SNR of 333 per scale-height ($\sim 60\ \rm{km}$). The shadow path traveled at 26.7~km/sec across Earth,
giving a linear resolution of 2.67~km/sample.  The 50-second central occultation measured Pluto's
atmospheric structure \citep{yfy08}.  However, the dataset also included close to three hours of additional data, from
two hours before the central event to one hour after.  We recently re-analyzed the entire dataset to search for
occultations by as-yet unknown rings, orbital debris, or satellites \citep{tfs11}. Results are shown in
Table~\ref{table:tau_aat} and Fig.~\ref{fig:tau}.

\begin{table}
\begin{center} 
\begin{tabular}{|l|l|}\hline
{\bf Search} & {\bf Upper Limit} \\ 
{\bf       } & {\bf (no detections)} \\ \hline \hline
Rings, 1500 km width & $\tau < 0.0035$ \\
Rings, 50 km width  & $\tau < 0.08$\\
Rings, 3 km width & $\tau < 0.035$ \\
Satellites along occultation path & $r$ $<$ 100~m \\ \hline 
\end{tabular}
\end{center}
\caption{Results of search for rings in 2006 Jun 12 AAT stellar occultation, from \citet{tfs11}.
\label{table:tau_aat}}
\end{table} 

\textbf{Other occultations.} Additional ring occultation results have been
reported by several groups.  \citet{mck08} and \citet{pbs06} report on
one additional dataset taken in Australia during the 2006 Jun 12 occultation. Their
limits are similar to ours.  \citet[][also Boissel in preparation 2011]{bsr08}
reports limits from four occultations in 2006-2008. Their limits are
also similar to ours. Neither of these studies found any dips attributable to
unknown satellites or orbital debris.

\section{Ring Upper Limits from HST Imaging}

The Hubble Space Telescope was used in June 2011\footnote{DD-12436, 5 orbits, PI Showalter} and June-July
2012\footnote{DD-12801, 34 orbits, PI Weaver} to search directly for rings around Pluto. These observations placed
Charon and Pluto in such a way as to facilitate subtraction of stray light. The reduced stray light allowed for a
much-improved lower limit to $I/F$ compared to previous results \citep[\eg][]{ss07}, where it was difficult to
distinguish between stray light and reflected ring light.  M. Showalter has presented preliminary results that showed a
limit on the normal $I/F < 3\times 10^{-7}$ outward of 12,000~km from Pluto. Inward of this distance the detectors were
saturated, so no limit could be calculated.

This new limit is 3-4 orders of magnitude better than the AAT occultation data, depending on the value of particle
albedo assumed. The AAT data is able to measure closer to the planet than HST, but for practical purposes the HST limit
is superior. 


\section{Impact Limits for New Horizons}

In this section I use the measured constraints on ring abundance to infer the population of mm-sized grains and thus the
risk to New Horizons on its path through the Pluto system.

An `absolute worst case' scenario can be examined if we assume that 100\% of the rings' optical depth is due to grains
of radius exactly $r_c$. In this case, Eqs. \ref{eq:iof} -- \ref{eq:ndamage} give us

\begin{equation}
N = {I\over F} {A\over{\mu_{sc}\, a\, \pi r^2 Q_{sca}}} 
\end{equation}


Plugging in the HST I/F limit and assuming an albedo $a = 0.05$, we find $N \approx 500$. This shows that observations alone,
when interpreted most conservatively, cannot technically rule out a danger to the spacecraft. However, in this case our
input assumptions are truly extreme enough that the computed value of $N$ is not really a useful statement about the
actual danger.  

In order to make a `plausible worst case' scenario, I make the more reasonable assumption that the grains are
distributed not unimodally but in a power law, where
\begin{equation} n(r)\, dr = r^{-q}\, dr \ .  
\label{eq:q} 
\end{equation}

The exponent $q$ indicates the slope of the distribution.  Values of $q < 3$ have most of their surface area in small
grains, while $q>3$ indicates a dominance of large grains.  Power laws are believed to be common in collisional systems
such as rings.  The actual distribution in rings across many orders of magnitude may well be much more complex, but it has
not been well measured.  Also, by choosing a wide range of power laws, we are likely to bound the actual
distribution.


I computed the $I/F$ for rings with $q$ in the range of 2--7. For each distribution, I calculated $Q_{sca}(r,\lambda)$
explicitly for every particle size.  All calculations were done at $\lambda = 0.5~\micron$. The size range extended from
$r=0.01\ \micron$ to $r=0.8\ \rm{mm}$ across 100 logarithmically spaced bins.  The upper size limit is consistent with
that used by \citep{tkk02}, and consistent with the drop-off in interplanetary impactors seen above 1~g by
\citet[][their Fig.~3]{gzf85}.  Neither the size limits nor the composition strongly affect the results.  I normalized
$n$ such that $I/F$ (Eq.~\ref{eq:iof}) matched that from HST.  It was then straightforward to calculate $N$ from
Eq.~\ref{eq:ndamage}. 

Because there is uncertainty in the true scattering law valid for ring particles \citep[\eg][]{te98}, I used two
end-members cases, with the expectation that the actual scattering properties should be bounded by these two.  The
first, Mie scattering, is valid for small, fresh, and/or spherical grains; here I assumed the particle composition to be
a silicate-ice mixture with index of refraction $1.33 + 0.001i$ (that is, relatively absorbing forward-scatterers). In
the second, I assumed Lambertian scattering with an albedo $a=0.05$ (that is, dark back-scatterers).  Following the
model of \citet{te98}, I have transitioned smoothly between these two scattering models at a size $r_{\hbox{trans}} =
300\ \micron$. More details on the scattering model are given in Appendix~A.


Results are shown in Fig.~\ref{fig:hst}. Each curve shows the number $N(r)$ of hits received as a function of particle
size $r$. The range in values for each curve (that is, the vertical height spanned by each curve) is due to the two
scattering models, with the Mie scattering model always being the higher of the two (that is, less reflective, and a
higher number density). The plot shows that for steep size distributions where $q\geq4$, NH is out of the `Danger'
zone, receiving $N < 0.1$ dangerous hits. However, for $q < 3.5$, the mission's safety cannot be assured.  Values of $q$
in the range $2-3$ yield $N \approx 2-20$ [\textit{Update: was $1-10$ in v1}] dangerous hits through the encounter.
Smaller values of $q$ result in larger $N$, because these distributions are dominated by larger grains.

The actual value for $q$ at Pluto is of course unknown. However, typical collisional ejecta has a $q_{ej}\approx3.5$ upon
initial creation \citep{dfs07}. Size-dependent processes in planetary rings almost always reduce $q$: Poynting-Robertson drag
and radiation pressure are both proportional to radius, reducing $q$ by 1.  Recent simulations of dust processes at
Pluto support $q = q_{ej}-1$ (Doug Hamilton, this workshop), suggesting that $q=2.5$ would not be unexpected for
dust at Pluto. Because the curves for $q=2-3$ are similar near near $r_c$, the inferred value for $N$ turns out to
be only weakly sensitive to changes to $q$. These low values of $q$ are the most dangerous ones and the most likely, and
thus \textit{the best observational limits cannot rule out a substantial impact hazard to the
spacecraft during encounter.}

In the Mie scattering case, I have assumed highly absorbing particles.  This is consistent with the dark particles
inferred for the Uranian ring, with $a \simeq 0.05$ \citep{cuz85}. Icy particles with a higher albedo are prevalent in
the outer solar system; Pluto and Charon's own albedos are in the range 0.3--0.5, suggesting the possibility of bright
ring particles. However, lacking any direct constraints on ring albedos, I use the most conservative observed values.
Brighter particles would decrease $N$.

\textbf{Future observations.} 2007 HST observations of the Pluto system \citep{ss07} placed limits of $I/F < 10^{-3}$ at
10,000 km. The limit was unable to go lower because of incomplete removal of stray light from Pluto-Charon in the
telescope optics. The recent 2011 HST observations improved the $I/F$ limit by some $3000\times$ by rolling the
spacecraft with Charon's orbital motion to allow for improved stray light removal.  2012 HST observations pushed this
limit closer inward in the Pluto-Charon system, but did not substantially change the numerical $I/F$ limit.  Future HST
observations (or other Earth-based observations) are unlikely to yield any significant improvement over the current
limit. Imaging in the infrared or mm (Spitzer, ALMA, JWST, Herschel) is much less useful than the optical because most
of the surface area is expected to be in small grains, where long wavelengths are less sensitive.

The $\tau$ limit from the occultation could be improved with additional observations using brighter stars and/or larger
telescopes. For instance, observing an occultation of an 11th magnitude star from a 10-m telescope would yield a signal
250$\times$ brighter the dataset used here, and an SNR some $15\times$ better. However, such an increase in sensitivity
would not be nearly sufficient to beat HST. Occultations will always be superior to HST for searching for small ring
arcs and isolated 100~m objects, and in the closest region toward the planet. In particular, appulses (where the shadow
path passes near the Earth but not over it) are more frequent than occultations, and would be quite valuable because
they often occur with brighter stars.

\begin{figure}
\centerline{\scalebox{0.6}{\includegraphics{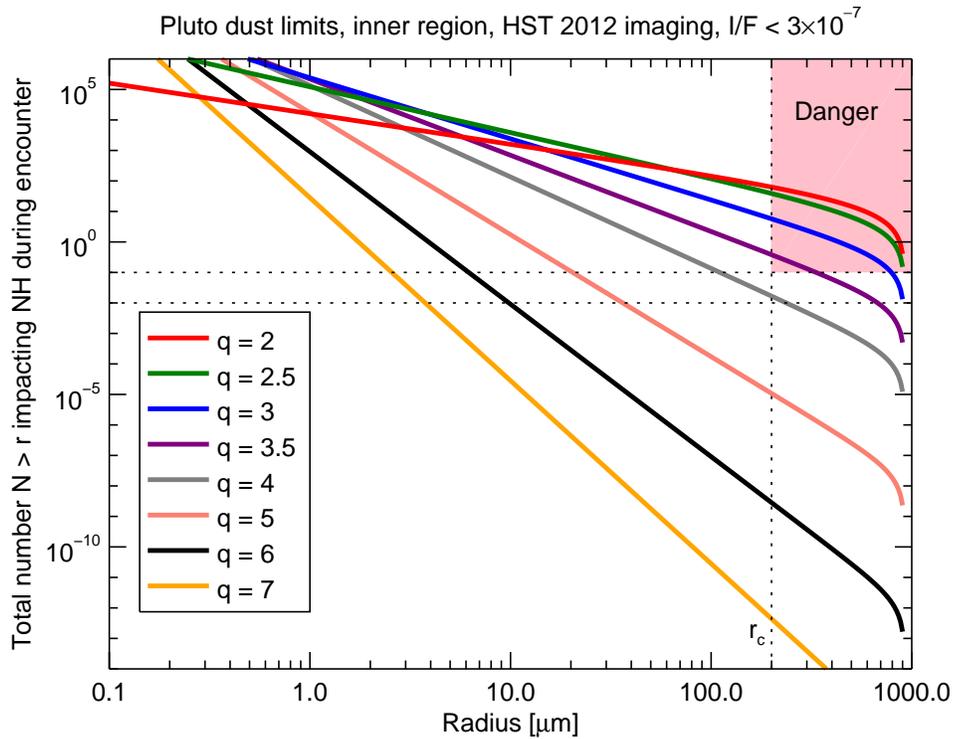}}}
\caption{Constraints on dust population from 2011-2012 HST observations. The curves indicate the different populations of
grains with different size distributions, using a combination of Mie and Lambertian scatterers. The visible $I/F$ of all
of the lines is identical and matches the HST-derived upper limit. The red quadrant indicates a population of particles
that violate the NH mission requirement of $N(r>r_c) < 0.1$. Models can place a limit of $N\ltsim20$ particles of size
$r_c$ (red curve).  \label{fig:hst}
} 
\end{figure}

\begin{figure}
\centerline{\scalebox{0.6}{\includegraphics{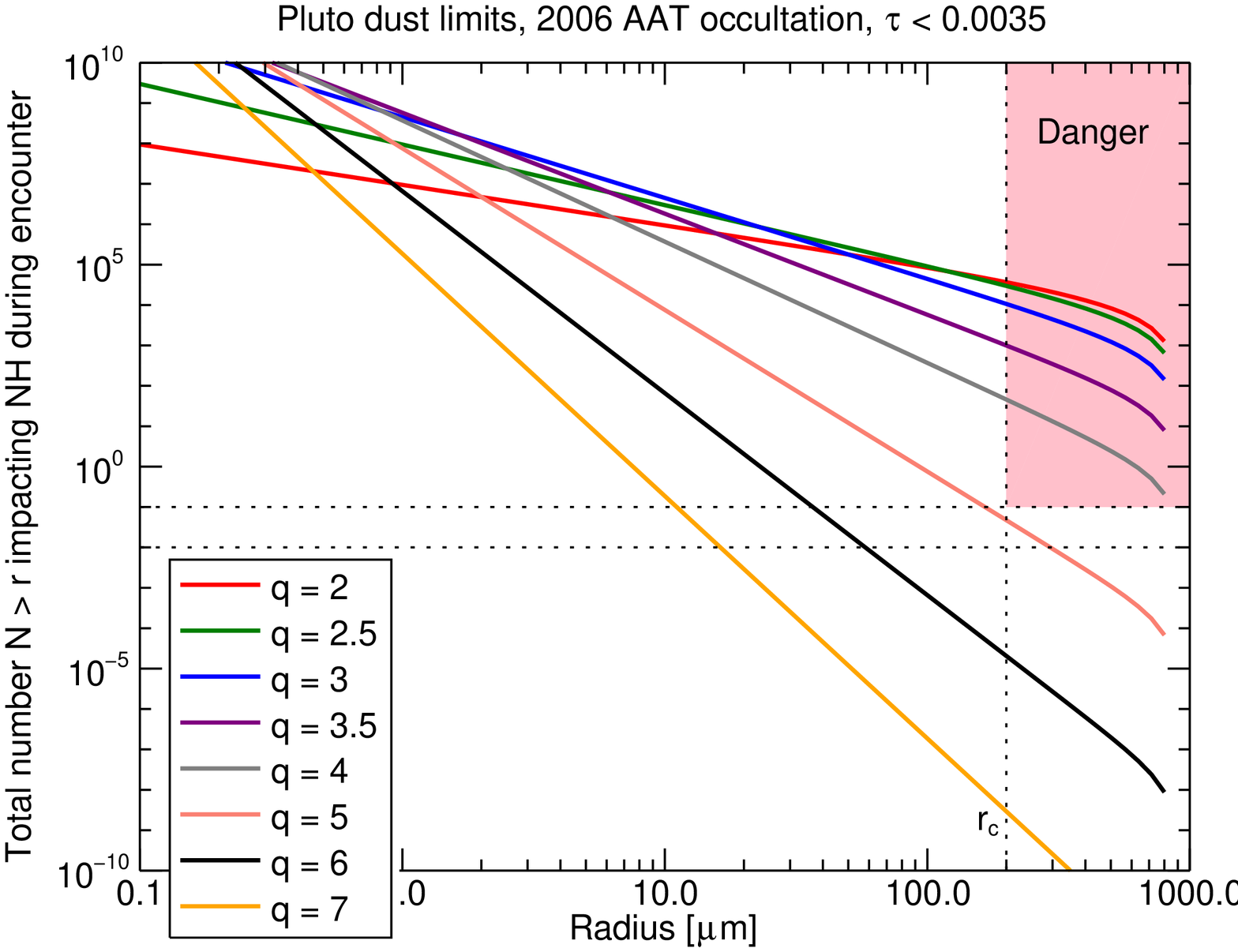}}}
\caption{Same as Fig.~\ref{fig:hst}, but using the limit on $\tau$ from the 2006 Jun 12 AAT stellar occultation. $N$ is
significantly higher here than with the HST figure, because of the weaker constraint on $\tau$.}
\label{fig:aat}
\end{figure}

\section{Acknowledgments}
Thanks to Leslie Young, Cathy Olkin, Mark Showalter, Hal Weaver, and Alan Stern for comments and suggestions. I also
thank Mark Showalter and Andrew Steffl for providing their unpublished data. Discussions with Matt Hedman were helpful,
and I am particularly grateful to him for noticing the error in how diffraction was handled in an earlier version of
this paper and checking the calculations in the current version.

\clearpage

\vfill \eject

\appendix
\section{Notes to version 2 -- December 2012}

In an earlier version of this paper, I assumed that to model the HST calculations, the scattering angle $\alpha$ was
close enough to zero that the single-scattering albedo $a$ could be written as

\begin{equation}
a = { Q_{sca}(r,\lambda) \over Q_{ext}(r,\lambda) } - 0.5
\label{eq:ssa}
\end{equation}
where the 0.5 is to account for the diffraction spike, which is formally part of the phase function but not seen in our
particular observations.

While this is an OK approximation for some particle sizes, it is technically incorrect, as can be easily verified at,
say, a value of $r = 0.01\ \micron$, which computes a negative $a$ given the refractive index and wavelength used here.
However, my code's actual computations conveniently avoided this issue by neglecting to include the 0.5 at all. As a
result, the plots in the paper over-estimated $a$ (that is, it erroneously computed that particles were bright, and easy
to detect), and thus under-estimated the impact rate $N$.  Matt Hedman pointed out this issue to me, and I have revised
the paper to address it.

This issue is very easy to fix. In this revised paper, I consider explicitly the value of the phase function $P(\alpha)$
for each particle, as shown in Eq.~\ref{eq:tau}. There is no reason not to include $P$; it was just an erroneous
simplification.

For small grains, the effect of including the phase function is minimal. For these, the phase function has little
structure near 0$\degrees$ which causes the singularities in Eq.~\ref{eq:ssa}, and the particles are naturally quite
reflective because no absorption happens due to their small size. As a result, the albedo approximation is accurate, and
the number of impacts $N$ is unchanged for the curves in Fig.~\ref{fig:hst} dominated by small grains ($q\ge3$).

For the large grains, resonances near backscatter causes structure in the phase function $P(\alpha)$, \textit{and} the
Mie particles naturally reflect less because they absorb more. For these grains, nearly all the light comes out in the
diffraction peak, with very little reflected. My original paper said in effect that these grains were easier to detect
than they really are. Handling the phase function properly causes a substantial increase in $N > r_c$ for distributions
dominated by large grains. In the case of $q=2$, this caused $N$ to be under-estimated by a factor of $\sim 100$.

\begin{figure}
\centerline{\scalebox{0.3}{\includegraphics{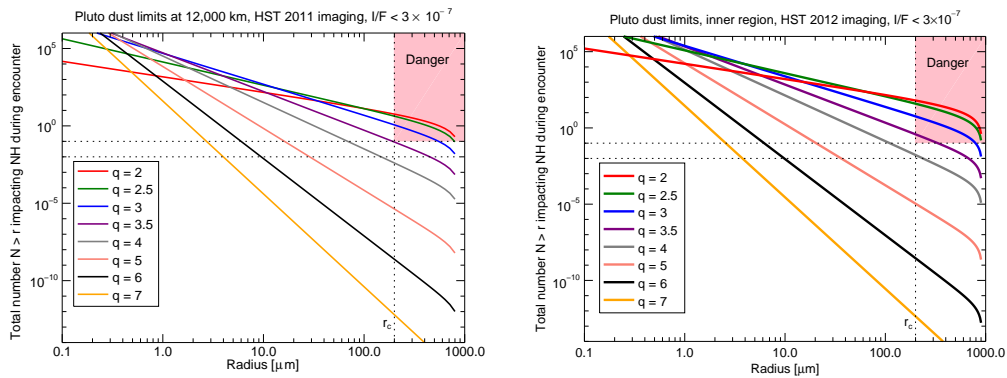}}
            \scalebox{0.3}{\includegraphics{pluto_dust_limits_hst_v2.eps}}}
\caption{Constraints on dust population from 2011-2012 HST observations. Left: Version 1, with Mie scattering only, and with
incorrect handling of diffraction. Right: Version 2, with Mie and Lambertian scattering (top and bottom, respectively),
and corrected diffraction handling. The difference is to increase $N$ by a factor of $\sim2\times$ for the curves that
enter the Danger region. This plot is identical to that in Fig.~\ref{fig:hst}, and is reproduced here for ease of comparison.}
\label{fig:hst_v1_v2}
\end{figure}

\begin{figure}
\centerline{\scalebox{0.6}{\includegraphics{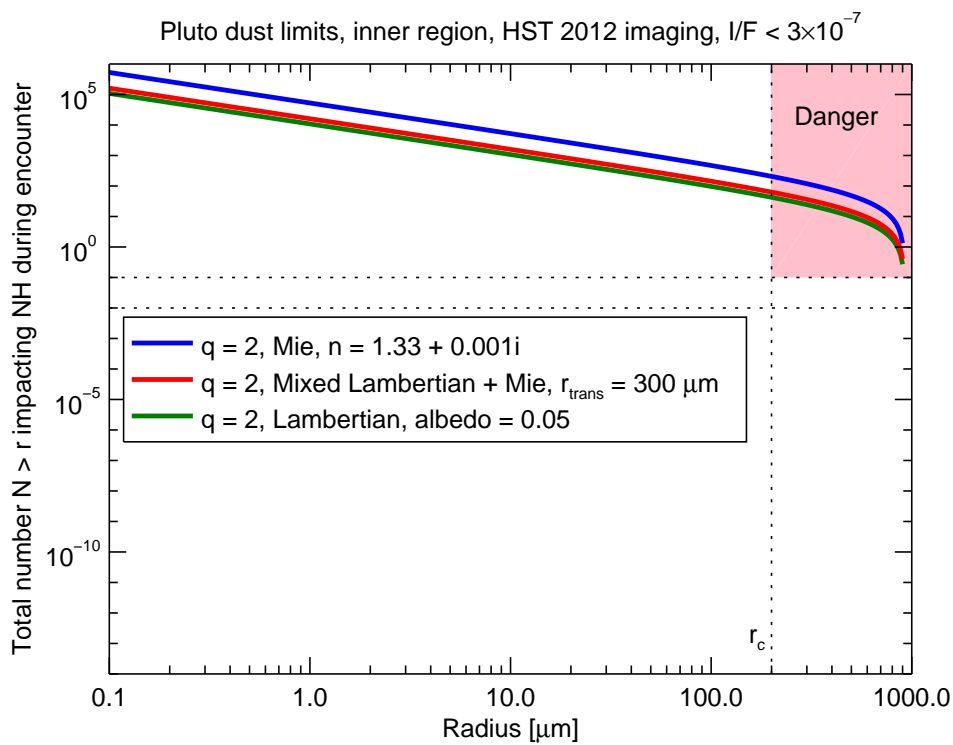}}}
\caption{Scattering curves for rings made of two different types of scatterers (Mie and Lambertian). The middle curve is
a combination of these two scattering models.  In the case of $q=2$, the ring is dominated by large grains and so the
Mixed phase curve lies close to the Lambert one. For $q>3$, the dominance of small grains causes the Mixed curve to lie
closer to the Mie phase curve. Any plausible scatterer should lie between the two end-member cases.}
\label{fig:hst_mie_vs_lambert}
\end{figure}

I made a second change to this paper, which was to address the fact that Mie scatterers may be a poor model in the first
place. While Mie is a commonly used scattering method for small grains, it becomes inaccurate for larger grains. Mie
methodology assumes that grains are perfectly spherical, homogeneous, and smooth, and that they are free of any cracks,
voids, inclusions, or surface roughness. Every single one of these terms will cause deviations from Mie, and they will
in general all cause grains to be more reflective (back-scattering). One physical example is a hailstone. Mie scattering
would calculate a large water ice sphere to be virtually invisible to any observers outside of the central diffraction
peak. But in reality hailstones (and by extension, an icy body such as Pluto, albedo $\sim$0.5) are extremely reflective
and easy to see. In these macroscopic bodies, scattering is due not to light waves traveling through the body (which Mie
describes), but by reflections by cracks and surface roughness. The size at which this transition from Mie to
macroscopic scattering occurs will vary based on the history and processing of the surface. Some experimental work has
shown it to be in the 10-100~\micron\ range \citep{sok87}, and a transition near this size was used to fit data from
Saturn's G ring \citep{te98}. In the current model, I chose a transition size of 300~\micron, in order to be somewhat
conservative, and hedge against the possibility that Pluto's rings might be caused by Enceladus- or Triton-like plumes,
which could produce a ring of fresh ice grains for which Mie scattering is a reasonable model.
Figure~\ref{fig:hst_mie_vs_lambert} shows a comparison between the different scattering models. 

For reference, the two versions of Fig.~\ref{fig:hst} are shown in Fig.~\ref{fig:hst_v1_v2}. These two changes partially
cancel each other out, but the overall effect is an increase in $N$ by a factor of $\sim 2$.

\subsection{Lambertian function}

Note that because the ring particles are not spatially resolved, the Lambertian phase function must be the
disk-integrated version, not the commonly used surface version. The disk-integrated phase function can be easily
derived from the surface one, and is taken as \citep[\eg][their Eq.~7]{pwr08}

\begin{equation}
P(\alpha) = {8\over{3\, \pi}} (\sin\alpha + (\pi - \alpha) \, \cos\alpha) .
\label{eq:p11_lambert}
\end{equation}

At backscatter ($\alpha \simeq 0\degrees$), this is just $8 / 3$. Note that Eq.~\ref{eq:p11_lambert} is
normalized such that

\begin{equation}
\int_0^{\pi} P(\alpha) d\alpha = 2.
\end{equation}

However, Eq.~\ref{eq:p11_lambert} includes only a reflectance term, and not a diffraction term. The value of
$P$ used in Eq.~\ref{eq:iof} should thus be half of this value, or $4 / 3$.

\clearpage

\bibliography{papers}




\end{document}